# Automating Spreadsheet Discovery & Risk Assessment

Eric Perry
Prodiance Corporation
5000 Executive Parkway, Suite 270
San Ramon, CA 94583 – USA
eric.perry@prodiance.com

**Abstract**

There have been many articles and mishaps published about the risks of uncontrolled spreadsheets in today's business environment, including non-compliance, operational risk, errors, and fraud all leading to significant loss events. Spreadsheets fall into the realm of end user developed applications and are often absent the proper safeguards and controls an IT organization would enforce for enterprise applications. There is also an overall lack of software programming discipline enforced in how spreadsheets are developed. However, before an organization can apply proper controls and discipline to critical spreadsheets, an accurate and living inventory of spreadsheets across the enterprise must be created, and all critical spreadsheets must be identified. As such, this paper proposes an automated approach to the initial stages of the spreadsheet management lifecycle - discovery, inventory and risk assessment. Without the use of technology, these phases are often treated as a one-off project. By leveraging technology, they become a sustainable business process.

**1.0 Spreadsheet Survey & Results**

During the course of 2007 and 2008, Prodiance Corporation and Jefferson Wells International hosted a monthly series of educational webcasts on Spreadsheet Remediation and Control. Thousands of senior finance and internal audit executives across a broad range of companies attended these events and responded to a series of survey questions. The questions and results of this spreadsheet survey are as follows:

- Q1: How important is it to have the proper safeguards and controls for your organization's mission critical spreadsheets?

Figure 1: Importance of Proper Spreadsheet Controls

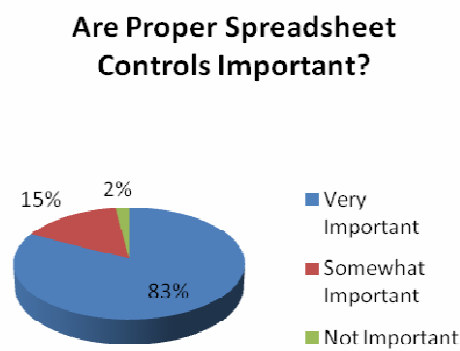

61

- Q2: Do you feel most organizations today have adequate spreadsheet controls in place?

Figure 2: Adequacy of Current Spreadsheet Controls

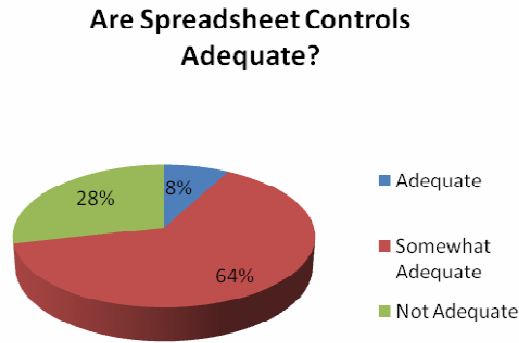

- 
- Q3: What is your organization currently doing about addressing spreadsheet controls?

Figure 3: Spreadsheet Controls Progress

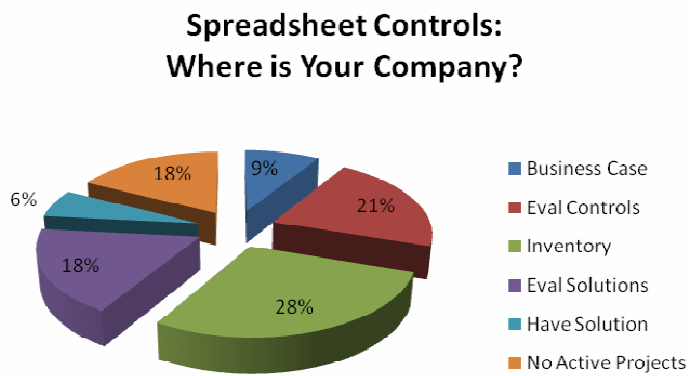

From the survey results, financial executives who responded felt having proper spreadsheet safeguards and controls in place is very important (83%), yet few felt that adequate controls were implemented in most organizations (8%). In addition, most organizations surveyed (76%) were in the early stages of implementing spreadsheet controls – either building a business case, evaluating existing controls, creating an inventory, or evaluating spreadsheet management and control solutions. These survey results confirm the need for spreadsheet discovery, inventory and risk assessment to further the adoption of spreadsheet controls.

**2.0 Spreadsheets – Which Ones Are Critical?**

In reading the latest research from industry experts [6] and analysts [7], we see that many organizations today are either unaware of the many potential risks uncontrolled spreadsheets can present to their organization, or too busy working on higher priority projects. Panko [5] asserts that "spreadsheet error rates are unacceptable in corporations today" and that the solution to addressing these high error rates and the risks they present requires comprehensive testing. However, many organizations may not be ready for testing because they are unaware of all of the spreadsheets that have been developed, where they are stored, and how to categorize them in terms of business risk. Prior to the recommended spreadsheet testing efforts, organizations first need to focus on spreadsheet discovery and the creation of a living inventory. A thorough risk assessment and categorization of this inventory will help determine which spreadsheets are mission critical, or pose high risk to an organization. Typically, high risk spreadsheets have a direct impact on the financial close and financial reporting process, tracking of P&L (e.g. revenue recognition), or regulatory reporting. In an earlier EuSpRIG paper, Croll [1] defined a critical spreadsheet as one where a "material error could compromise a government, a regulator, a financial market, or other significant public entity and cause a breach of the law and/or individual or collective fiduciary duty." Often a good place to start looking for critical spreadsheets is the financial close process.

**3.0 Discovery & Risk Assessment – Typical Approach**

Unless spreadsheet use within an organization is isolated to a single finance department, discovery, inventory and risk assessment can be a challenging exercise. In many cases, thousands or even hundreds of thousands of spreadsheets exist across a distributed network and multiple geographic locations. End users often have no idea where latest the versions of spreadsheets are stored or where prior versions are archived. As such, creating an inventory is often a manual and time and resource intensive effort. Once an inventory is created, it is quickly outdated as new spreadsheets are often created on a daily, weekly and monthly basis. Spreadsheet discovery is an ongoing challenge, and industry experience has shown the initial process can absorb countless man hours and extend for up to 2 or 3 months before an accurate inventory is created.

In addition, once critical spreadsheets are discovered, those creating the inventory (typically internal auditors) often have limited knowledge of spreadsheet use and complexity. This makes risk assessment difficult, and without a quantifiable methodology to assess risk and impact to business, it can be impossible. Many organizations have compiled risk matrices, but applying them manually can lead to inconsistent results.

**4.0 Discovery & Spreadsheet Risk Assessment - Automated Approach**

Automating spreadsheet discovery requires the application of software technology and a proven methodology. Leading audit firms [2] recommend using "…commercially available or homegrown tools that can be configured to scan network resources and return a list of all spreadsheets used in the organization. Providing that all relevant resources are scanned, this technique will result in the most complete spreadsheet population list possible."

The following guidelines have been used recently and with a great deal of success at several leading banks and insurance companies.

- First, any and all computers should be identified; including corporate file shares, document and records management repositories, and employee PCs. Particular attention should be given to those suspected of containing critical spreadsheets.
- Then, these computers should be scanned initially and on a scheduled basis (i.e. weekly is recommended) to create a centralized inventory.
- Scanning criteria should include any and all files containing file names known to be critical (e.g. revenue recognition.xls, 2Q_2008_earnings.xls, etc.), those that have been created or last saved/modified during the financial close cycle, and those created or modified since the last discovery.
- The software should also have options to capture any spreadsheets with incorrect or missing file extensions, or those compressed in ZIP folders.

- Advanced options for finding any linked (e.g. dependent spreadsheets or data sources) should strongly be considered. If a spreadsheet is deemed critical, then any spreadsheets feeding data into it should also be considered critical.
- Finally, read only permission must be granted to any user or software discovering files on employee desktops, or an optional agent should be deployed to run undetected by the local user (and perform the scan).

**5.0 Spreadsheet Risk Assessment Methodology**

Initial discovery results are likely to contain an inventory of thousands of spreadsheets, not all of which are critical. To efficiently identify the critical spreadsheets in your inventory, leading audit firms have prescribed an evaluation of magnitude (or materiality) and complexity [2]. Spreadsheet materiality can be evaluated by analyzing the following criteria: cell values, currency values, operational values, document properties, file names, sheet names, file path, external links. Consider the following example to define spreadsheet materiality:

Figure 4: Spreadsheet Materiality Criteria

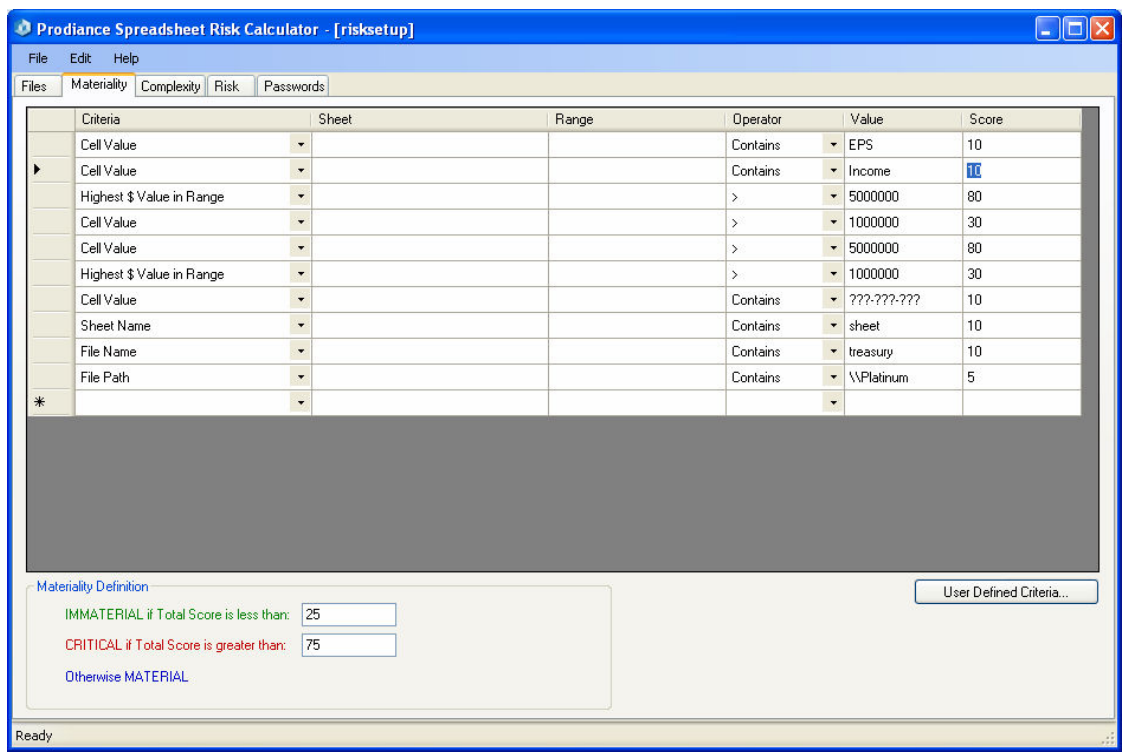

For example, using the grid in Figure 4, we can assign a score of "10" to any spreadsheet with a cell value containing the text string "Income", and an additional "80" points (for a total of 90 points) to any spreadsheet with a $ value exceeding "5,000,000". A spreadsheet then satisfying both of these criteria would be classified as "CRITICAL" according to the Materiality Definition.

Similarly, spreadsheet complexity can be evaluated by analyzing key spreadsheet elements, including: worksheets, formulas, formula errors, array formulas, nested IFs and number of levels, external links, macros, named items, invisible cells, hidden sheets/rows/columns, very hidden sheets, workbook size, password protection, and workbook size. Consider the following example to define spreadsheet complexity:

Figure 5: Spreadsheet Complexity Criteria

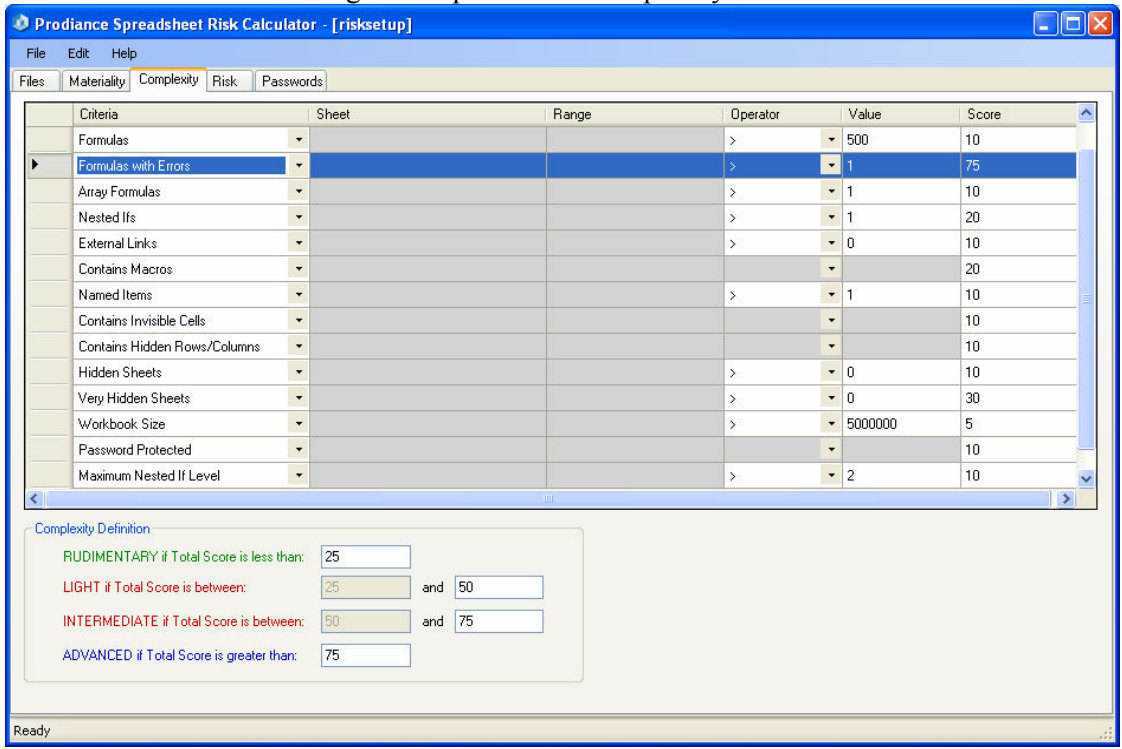

For example, using the grid in Figure 5, we can assign a score of "75" to any spreadsheet containing more than "1" formula error, "10" more points to any spreadsheet that "Contains Invisible Cells" and an additional "10" points to any spreadsheet that is "Password Protected". A spreadsheet meeting all 3 of these complexity criteria would score a "95" points and rate "ADVANCED" in terms of complexity.

The final step in automating risk assessment is to assign a risk level. The following table is an example of how spreadsheet risk can be assigned based on the intersection of materiality and complexity.

Figure 6: Spreadsheet Risk Matrix

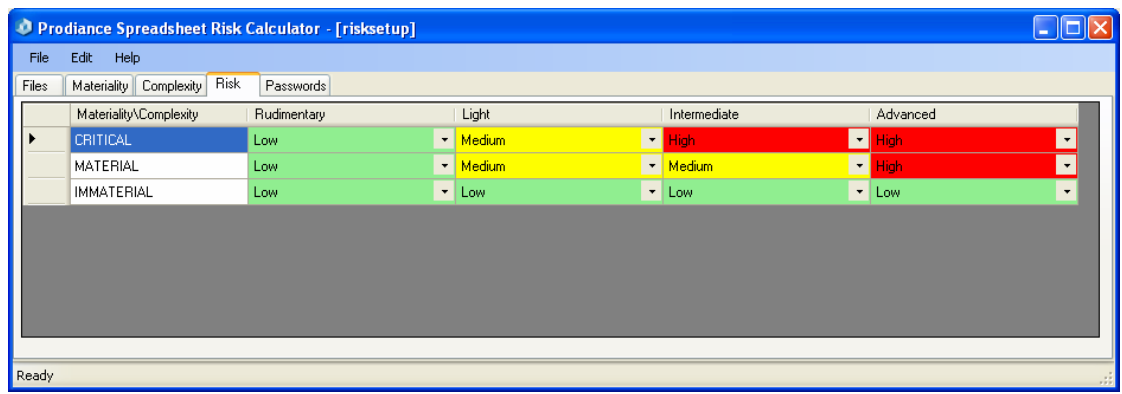

For example, a spreadsheet that ranks "CRITICAL" for Materiality and "Intermediate" for Complexity would rank as "High" Risk according to the grid in Figure 6.

However, to fully automate spreadsheet discovery and risk assessment, a software application should execute the aforementioned process in an automated and scheduled process as follows:

1. Discover all relevant spreadsheets across the network
2. Create centralized inventory
3. Perform risk assessment based on pre-defined materiality and complexity criteria
4. Generate and distribute initial spreadsheet inventory and risk report
5. Repeat the entire process per a weekly or monthly schedule to identify any new high risk spreadsheets

**6.0 Conclusion**

Hoye concluded that "organizations may benefit from software solutions that deliver real-time monitoring of critical spreadsheet activity, providing management with transparency into the control process" [4]. Real world experience has proven that leveraging software technology provides significant advantages to help overcome the challenges of the typical approach described above, including:

- Reducing the 2-3 month timeframe of the typical (manual) approach down to 2-3 days
- Conducting a comprehensive scan of an entire IT network for any and all spreadsheets in existence, including corporate file servers, content repositories, and even employee PCs
- Managing a centralized, live inventory of all spreadsheets present across an organization
- Providing an automated risk assessment framework and methodology (that is consistent with auditor guidance) to help categorize spreadsheets according to risk level
- Enabling discovery to run as a continuous process to help identify any newly created, high risk spreadsheets, ensuring the centralized inventory is always current
- Providing management and auditors with visibility into the discovery and risk assessment process via automated reports (delivered via email) of the inventory and any high risk spreadsheets or control policy violations

Typical approaches to spreadsheet discovery and risk assessment resulted in one-off projects with inconsistent, incomplete and results that are quickly outdated. By leveraging technology and best practices, spreadsheet discovery and risk assessment reaches maturity as a sustainable and automated business process.